\documentclass[11pt,a4paper]{revtex4-1}

\usepackage[british]{babel}
\usepackage[utf8]{inputenc}
\usepackage{graphicx}
\usepackage{hyperref}
\usepackage{epsfig}
\usepackage{natbib}
\usepackage{xspace}
\usepackage{xcolor}
\usepackage{orcidlink}
\usepackage{mathrsfs}
\usepackage{amsfonts}
\usepackage{amsmath}
\usepackage[normalem]{ulem}
\usepackage{hyperref}
\usepackage{enumitem}
\usepackage[title]{appendix}

\usepackage{soul} 

\begin{document}

\title{Inequality in a model of capitalist economy}

\author{Jhordan Silveira Borba}  
\affiliation{Instituto de Física, Universidade Federal do Rio Grande do Sul, 
Av. Bento Gonçalves 9500, 90560-001, Porto Alegre (RS), Brazil}

\author{Sebastian Gonçalves}
\affiliation{Instituto de Física, Universidade Federal do Rio Grande do Sul, 
Av. Bento Gonçalves 9500, 90560-001, Porto Alegre (RS), Brazil}

\author{Celia Anteneodo}
\affiliation{Department of Physics, PUC-Rio, Rua Marqu\^es de S\~ao Vicente  225, 22451-900 Gávea, Rio de Janeiro (RJ), Brazil}
\affiliation{National Institute of Science and Technology for Complex Systems, 22290-180, Rio de Janeiro (RJ), Brazil}


\begin{abstract}
We analyze inequality aspects of the agent-based model of capitalist economy named {\it Social Architecture of Capitalism} that has been introduced by Ian Wright. The model contemplates two main types of agents, workers and capitalists, which can also be unemployed. Starting from a state where all agents are unemployed and possess the same initial wealth,  the system, governed by a few simple rules, quickly self-organizes into two classes. After a transient, the model reproduces the statistics of many relevant macroeconomic quantities of real economies worldwide, notably the two regimes of the distributions of wealth and income.  
We perform extensive simulations testing the role of the model parameters (number of agents, total wealth, and salary range) on the resulting distribution of wealth and income, the social distribution of agents, and other stylized facts of the dynamics. 
Our main finding is that, according to the model, in an economy where total wealth is conserved and with a fixed average wage, 
the increase in wealth per capita comes with more inequality.


\end{abstract}

\maketitle
 
\section{Introduction}
\label{sec:introduction}

The search of a first-principle model capable of elucidating the emergence and robustness of Pareto's law~\cite{paretonovo} is a central challenge in the field of Econophysics. 
Actually, as originally reported by Pareto, income and wealth real-world  present two regimes: the upper tail follows a power-law (Pareto's law), while the lower values display exponential  behavior ~\cite{dragulescu2001,yakovenko2023}.
Such  first-principle model would provide a fundamental understanding of the mechanisms governing these properties, transcending mere empirical observations to reveal the possible underlying principles that drive economic inequality and its evolution over time.
In this regard, agent-based models~\cite{chakraborti2011}, 
particularly kinetic exchange models~\cite{metzler2023}, allow for the simulation of transactions among agents ---such as individuals, firms, or other economic entities--- that drive economic dynamics and help identify the underlying processes shaping these systems. 
This approach has paved the way for numerous efforts over the past few decades to explain a variety of phenomena, particularly the unequal distribution of wealth (see, for instance, ~\cite{boghosian2014a,metzler2019,chakraborti2000,dragulescu2000,bouchaud2000} and \cite{yakovenko2009,chakraborti2011,chakrabarti2013} for reviews).  
It is well known that unbiased random exchanges result in a condensed phase populated by the poorest agents~\cite{boghosian2014a, Cardoso2023}. To prevent such condensation and generate realistic distributions, extra mechanisms or regulatory controls have been introduced. These mechanisms can include the introduction of specific tax structures~\cite{LIMA2022,Calvelli2023,Dias2024} or other wealth constraints, which, while effective in shaping outcomes, can make the results less purely emergent and more dependent on externally imposed rules.
Let us give some examples.
In random asset exchange models, the  introduction of savings can mitigate condensation~\cite{chakraborti2000}, stabilizing the wealth distribution with a most probable value away from zero.
The 
two-regime wealth/income distributions can be reproduced by the introduction of disorder in saving propensity~\cite{ Chatterjee2004}, and also by wealth-based advantage mechanisms~\cite{Calvelli2023,Vallejos2018}.  
When networks are considered instead of the more common mean-field setup (where all agents are potentially connected), topology can influence the wealth distribution~\cite{KOHLRAUSCH2024}.
Moreover, the power-law exponent can be altered by diverse factors, for instance by the distribution of saving propensities~\cite{AYDINER2018}, or by network position exchanges~\cite{metzler2019}. 
In the present paper, using agent based simulations,  we investigate a model that,  simultaneously gives rise to two key phenomena: the emergence of the dual-regime structure in wealth and income distributions, alongside the development of a two-class society. 

In his seminal work ``The Social Architecture of Capitalism'' \cite{WRIGHT2005589},  Ian Wright explores the intricate dynamics that results in the self-organization of a capitalist society with a given distribution of wealth among its members. Using an agent-based model known as Social Architecture (SA) model, Wright provides a theoretical framework aiming to explain the ubiquitous presence of Pareto's law, which reflects that a small percentage of a population controls a large portion of a particular resource such as wealth or income. Wright's work offers insights into how individual interactions and institutional structures collectively shape economic distributions.  
The model effectively reproduces the so called, ``80/20 rule''~\footnote{
 The $80/20$ rule derives its name from Pareto's observation of the Italian economy, where he found that roughly 20\% of the population owned $\sim 80$\% of the resources. The term is used generically, although the ratio $f:(1-f)$ may vary across different countries and societies, reflecting a broader pattern in which a small proportion of inputs or entities accounts for a large share of outcomes or resources~\cite{boghosian2014,boghosian2014a}.}, but, most notably, it naturally generates the two-pattern shape of the wealth or income distributions (exponential-like, for low wealth/income, power-law for high wealth/income). 

Wright compares his results with real data, demonstrating that the model captures a significant portion of the dynamics of the capitalist system. 
In addition to the two regimes of wealth and income distributions, the model can reproduce the statistics of other economic indicators, such as
firm size (power law), firm size growth rate (double exponential), firm demises per month (log-normal), and recession duration (exponential). These patterns emerge naturally and robustly from the simple set of rules that define the model. 
Hence, the SA model lays the foundation for exploring various economic phenomena, starting from microscopic (or agent-based) dynamics controlled by a few parameters. Several follow-up studies have been published~\cite{Cottrell2009,WRIGHT2010,LAVICKA2010,ISAAC2019,VILLA2022,MAX2024}, but the SA model itself has not been thoroughly analyzed.

Our present goal is to investigate aspects related to inequality in income and wealth distributions using the SA model. We aim to conduct a comprehensive examination of the role played by the model parameters as well as certain details of Wright's original model. Specifically, we seek to answer the following questions: How do the distributions of income and wealth across different classes depend on parameters such as system size, total wealth, and average wage? How robust are these results? Are they universal and, if so, to what extent? Or are they sensitive to key parameters, and if so, to what degree?

By analyzing the social and economic structures that contribute to income disparities and wealth concentration in capitalist societies, we aim to identify the key parameters and dynamic rules that give rise to Pareto-like distributions and other significant economic measurements. Ultimately, this paper seeks to contribute to a broader understanding of inequality and concentration phenomena in socio-economic systems.

\section{Model}

The model implemented in this paper is the original SA proposed by Ian Wright~\cite{WRIGHT2005589}. However, we have chosen a slightly different notation and terminology to align with those commonly used in econophysics and agent-based modeling.
The society consists of $N$ agents (which are labeled $i=1,\dots,N$ and are not necessarily individuals, but can represent other economic entities), where each agent $i$ has a positive integer quantity $w_i(t)$ representing wealth, 
which varies over time. 
Here wealth is defined as an amount of money, assets or goods that can be eventually transformed into money, owned by each person in a population. Although this quantity does not need to be an integer, we follow the original proposal, as this distinction is not critical to our analysis.
The size of the population, $N$, and the total wealth of the system, $W= \sum_{i=1}^N w_i(t)$, are conserved, so the wealth per capita $\overline{w}=W/N$ is constant. 
Agents can be in one of three classes: employees (working class), employers (capitalist class), or unemployed. 
Then, each agent is characterized by an index $e_i\neq i$, which identifies their employer: $e_i = j$ if agent $j$ is the employer of agent $i$, and $e_i = 0$ if agent $i$ is unemployed or employer.
Therefore, at any time $t$,  the state of the entire economy is defined by the set of pairs $S(t) = \left\{ \left( w_i(t), e_i(t) \right) :1\leq i\leq N \right\}$.
A firm consists of a set of employees and an employer, which is the only owner of the firm.

The model has four parameters: system size $N$, wealth per capita $\overline{w}$, and the minimum ($p_a$) and maximum ($p_b$) wages employees can receive. Unless stated otherwise, all random selections (agents, wealth aliquots, wages, etc.) are uniform.

 Besides agents (employees and employers), the model includes a market value $V$, representing goods and services produced by firms. $V$ grows with consumer expenditures and decreases with firm revenues. Firms compete for a share of this pool rather than receiving direct payments. 

In the initial state, all agents have the same wealth ($w_i(0)={W}/N$) and are all unemployed ($e_i(0)=0$).
At each time step (which we set to be a month as in the original paper), the following six steps are executed $N$ times, to give each agent the opportunity to be active once per month on average.

\begin{enumerate}[leftmargin=*]
\item  
{\bf Agent selection} --  
An agent $i$ is randomly selected. 
Then, the following rules are applied, being 3, 4 and 6 related to wealth exchanges, while rules 2 and 5 are related to changes of status.  
\item {\bf Hiring} -- Only if $i$ is unemployed, then:
\begin{enumerate}[leftmargin=*]
 \item A potential employer $j$ is selected from the set $H$  of all agents except the employees, with probability  $P (j)= w_{j}/\sum_{n\in H}w_{n}$. 
 
\item   If $ w_j \ge \overline{p}$ (average wage, or salary, from the distribution of wages), then: 
the agent $i$ is hired by $j$.   
If $j$ was unemployed, it becomes hence an employer.

\end{enumerate}
\item  
{\bf  Expenditure} (on goods and services) --
A random integer $w$ in the range  $\left[0,w_{k}\right]$  is selected for a random agent  $k\neq i$. This amount, representing the agent's expenditure, is subtracted from their wealth and added to the market value, so:
\begin{eqnarray} \notag
w_k  &\to& w_k-w, \\ \notag 
V &\to& V+w.
\end{eqnarray}

\item 
{\bf Market revenue} (from sales of goods and services) -- Only if $i$  is not unemployed, a random integer $w \in [0, V]$ is  transferred from the market value to the agent $i$:   
\begin{eqnarray}   \notag
   && V \to V - w, \\  \notag
&& \begin{cases} 
  w_i \to w_i + w,         &\mbox{if  $i$ is an employer,}\\
  w_{e_i} \to w_{e_i} + w, &\mbox{if  $i$ is an employee.} 
\end{cases}
\end{eqnarray}
 In all cases,  $w$ counts as revenue for the firm owner, either if $i$ is the owner itself or if $i$ is an employee (in which case $e_i$ is the owner).

\item {\bf Firing --} Only if \textit{i} is an employer, then:
 
\begin{enumerate}[leftmargin=*]
    \item The number $m$ of employees to be fired is defined according to the  formula 
$ m=max(n_{i}-\frac{w_{i}}{\overline{p}},0)$, 
where $n_i$ is the number of agents employed by agent $i$.

\item A number $m$ of agents from the list of $i$'s employees are randomly selected to be fired.

\item
Furthermore, if all workers are fired, then the employer $i$ becomesunemployed.

\end{enumerate}
\item 
{\bf Wage payment --} Apply only when agent $i$ is an employer. 
For each agent $j$ employed by $i$, the following transfer occurs  
\begin{center}  
  $w_i  \to w_i-p$,  \\
  $w_j  \to w_j+p$,
\end{center}
where $p$ is a discrete amount randomly selected from the interval $[p_a,p_b]$, but if $p > w_i$, then a new wage $p$ is selected from the interval $[0,w_i]$.

\end{enumerate}

The update is asynchronous, that is, the system state is updated each time each rule is applied. 
At the end of $N$ iterations of these six rules, $t$ is incremented by one unit (i.e., one month). 
Unless otherwise stated, the values of the parameters used in the simulations of the model are: system size $N = 1000$ agents, wealth per capita $\overline{w} = 100$, salary range $[p_a,p_b]=[10,90]$. Typically, the maximum simulation time is $t_{max}=1000$ years, and the values of all state variables are recorded at the end of each year. 
For statistical calculations, data were recorded after reaching a steady state, discarding a short transient period.

\section{Preliminary  analysis}
\begin{figure}[h!]  
    \centering
    \includegraphics[width = 0.6\textwidth]{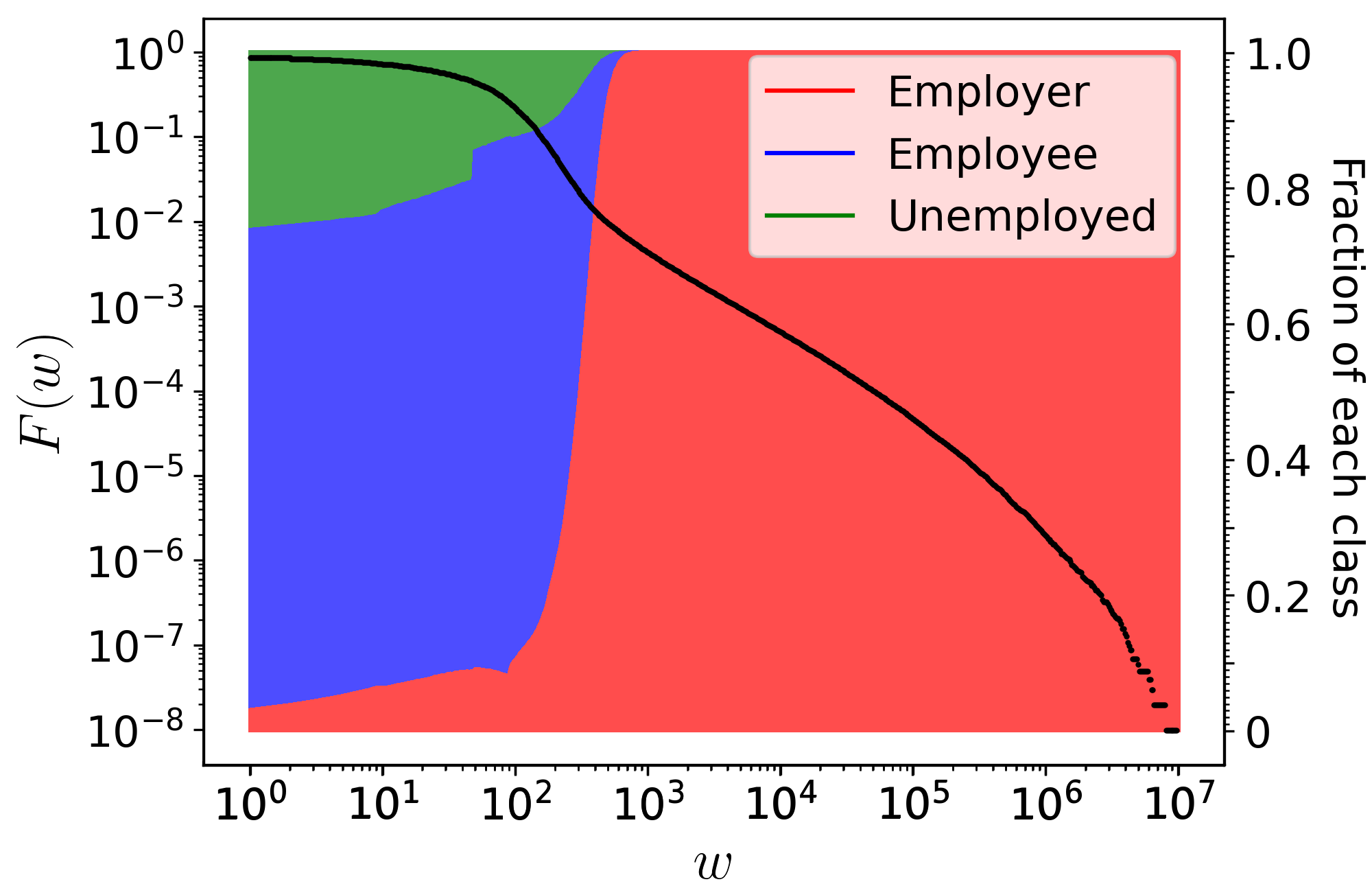}
    \caption{Complementary cumulative wealth distribution (black symbols) $F(w)$ and fraction of the population in each class (color plot) vs. wealth $w$. Result for  $N\simeq 10^5$, $\overline{w}=100$ and wage range $[10,90]$. 
    Data accumulated over 1000 years, recorded at the end of each year for each agent. 
    }
    \label{fig:portrait}
\end{figure}

The distribution of the population among the three classes of agents, along with the wealth distribution within each group, which emerges from the dynamics of the model, is displayed in Fig.~\ref{fig:portrait}. 
We observe a significant concentration of wealth within the capitalist class; however, it does not exhibit the typical pattern of wealth inequality ---where most agents have no wealth--- commonly found in binary exchange models from the existing literature when there is no intervention (such as taxes or assistance programs)~\cite{moukarzel2007,boghosian2014,CARDOSO2020,LIMA2022}.

\begin{figure}[h!]     
\includegraphics[width = 0.45\textwidth]{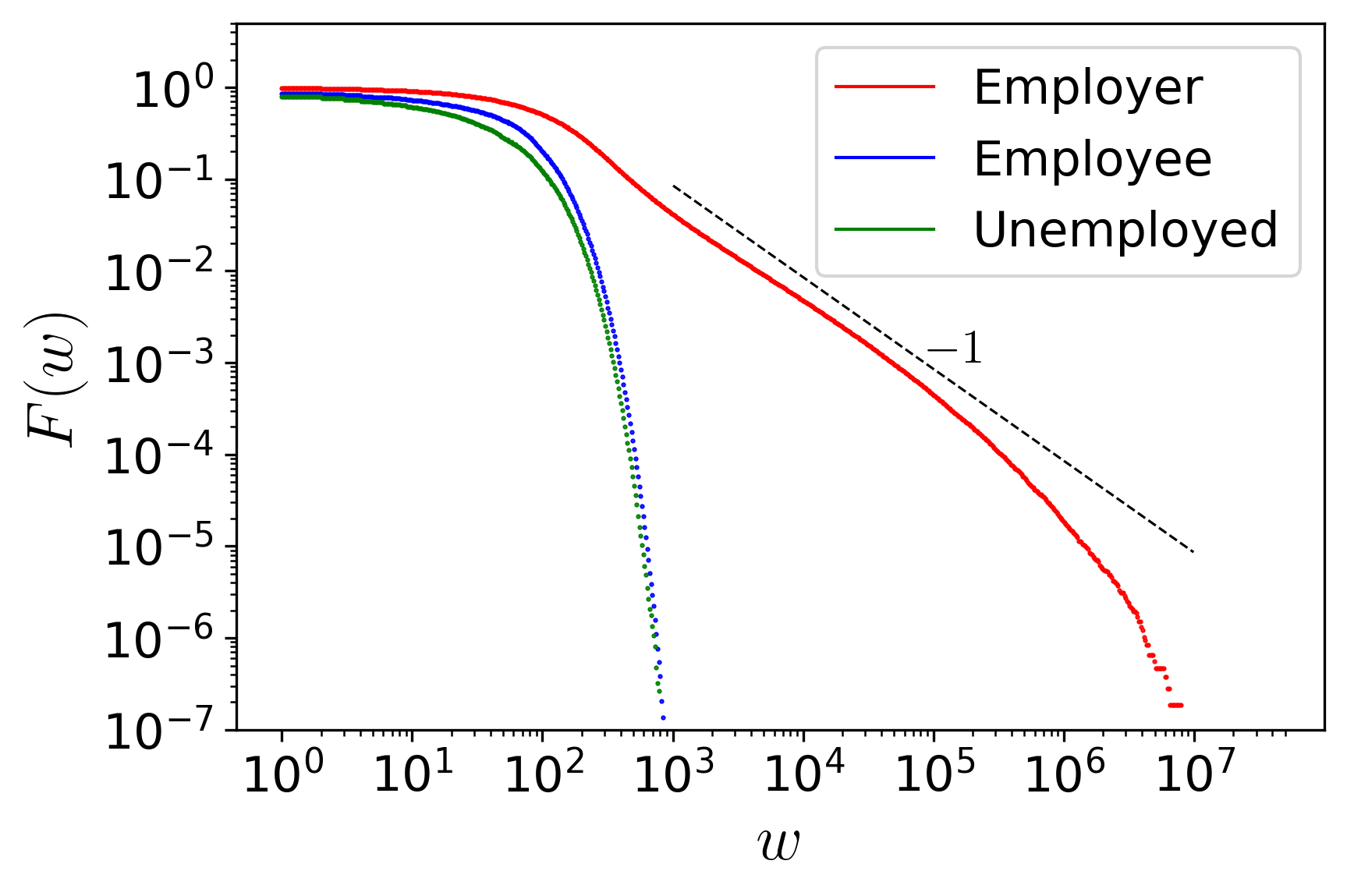} 
\caption{Complementary cumulative distributions of the wealth per capita within each class.  We used $N=102400$,  $\overline{w}=100$, and salary range $[10,90]$. The segment with slope -1,  which represents an upper bound, is drawn for comparison.
Data of a single realization were accumulated over  1000 years.
}
\label{fig:wealth-histograms}
\end{figure}

The separate distributions of wealth per capita within is class is shown in  Fig.~\ref{fig:wealth-histograms}, exhibiting a long tail for the capitalist class, with workers and unemployed present an abrupt cut-off. 
This decomposition helps to understand the origins of the two regimes observed in the global distribution of wealth, as shown in  Fig.~\ref{fig:portrait}.

In the following sections we will take a closer look on this impact of system size and wealth per capita on the economy dynamics. 
We will also see that a key parameter of the model is $R\equiv\overline{w}/\overline{p}$.  
In fact, first note that, according to the rules, the dynamics will not be impacted if wealth is scaled by an arbitrary quantity, for example, by $\overline{p}$, which only alters the monetary unit. 
Typically, we will  use  $ \overline{w}=100$ and $\overline{p}=50$, unless indicated differently.  
Given the chosen initial state, where agents start as unemployed with the same wealth $\overline{w}$, then, $R\ge 1$ is necessary to comply with   $w_j \geq\overline{p}$ (rule 2(b)), to start hiring, and to  prevent the system from becoming trapped in its initial state.

The remainder of the paper is organized as follows: In Sec.~\ref{sec:wealth},  we investigate the wealth distribution in more detail, focusing systematically on the effects of system size $N$, wealth per capita $\overline{w}$ and 
wage interval. 
In Sec.~\ref{sec:income}, we conduct a similar analysis for the income distribution. Finally, in Sec.~\ref{sec:final}, we present a discussion and the main conclusions.

\section{Wealth distribution}
\label{sec:wealth}

In this section, we analyze the wealth distribution by varying key model parameters: system size, total wealth, and average wage.
 
\subsection{Influence of system size $N$}

\begin{figure}[h!]   
\includegraphics[width =0.45\textwidth]{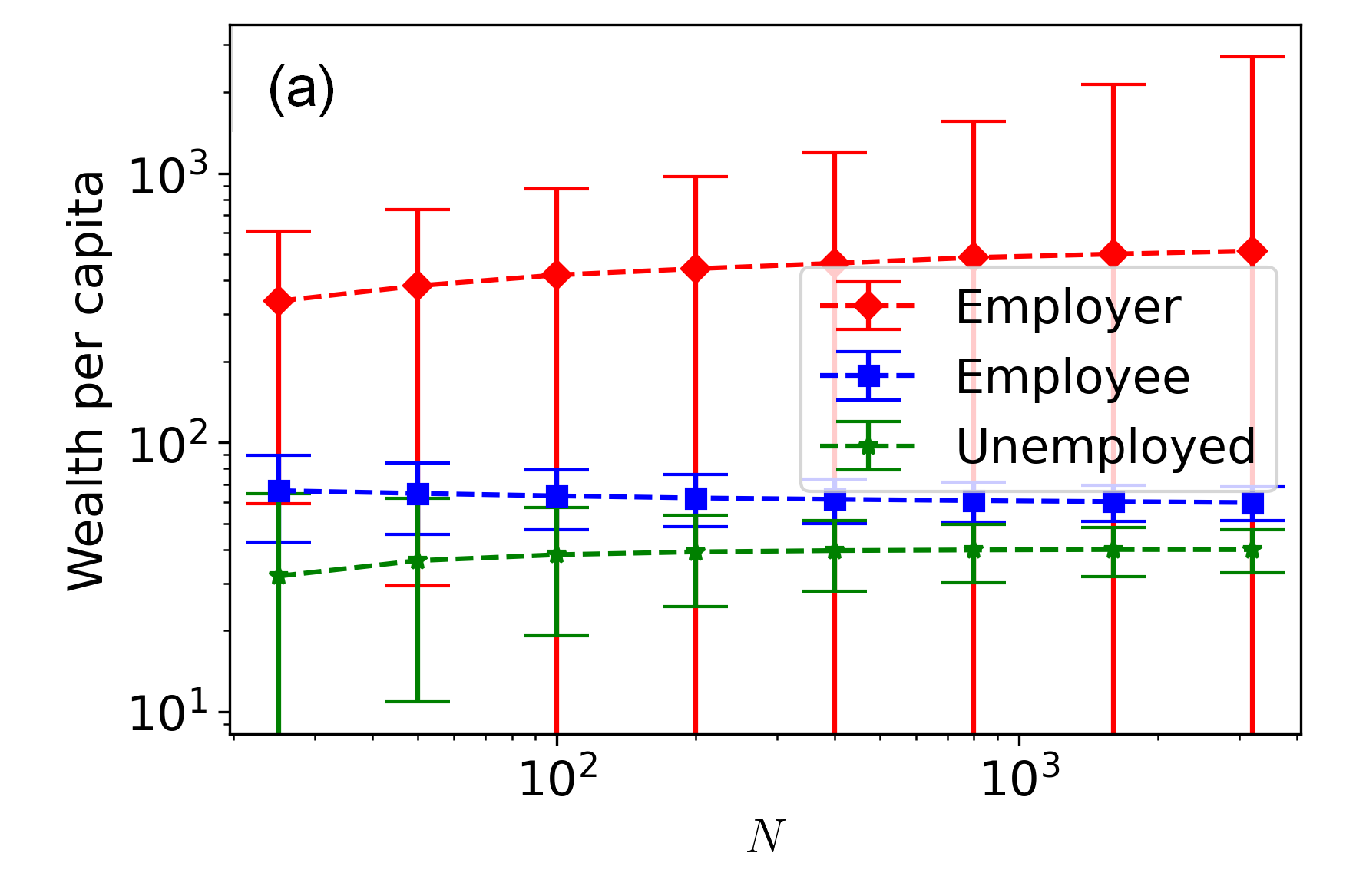} 
\includegraphics[width = 0.45\textwidth]{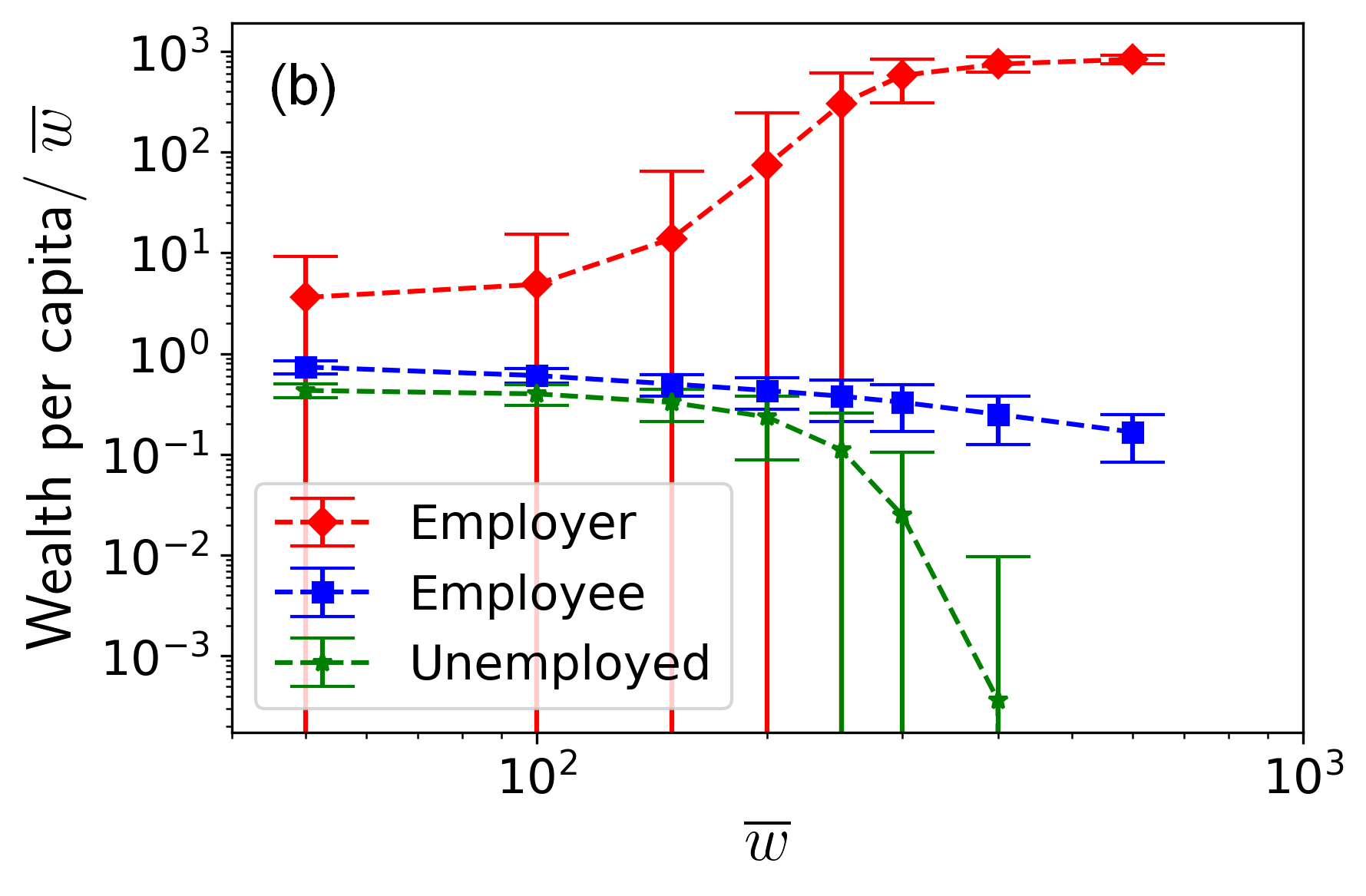}
    \caption{ 
 Wealth per capita within each class vs (a) $N$, for $\overline{w}=100$, and (b) $\overline{w}$, for $N=1000$ (b). 
    For each abscissa value, we plot the average (symbols) and standard deviation (bars) computed over  100 realizations of 1000 years. 
    In all cases, the wage range is [10,90]. The missing points for unemployed in (b) is due to the disappearance of this class. 
    The dashed lines are a guide to the eye.
}
    \label{fig:N&w}
\end{figure}

\begin{figure}[h!]    
    \includegraphics[width = 0.45\textwidth]{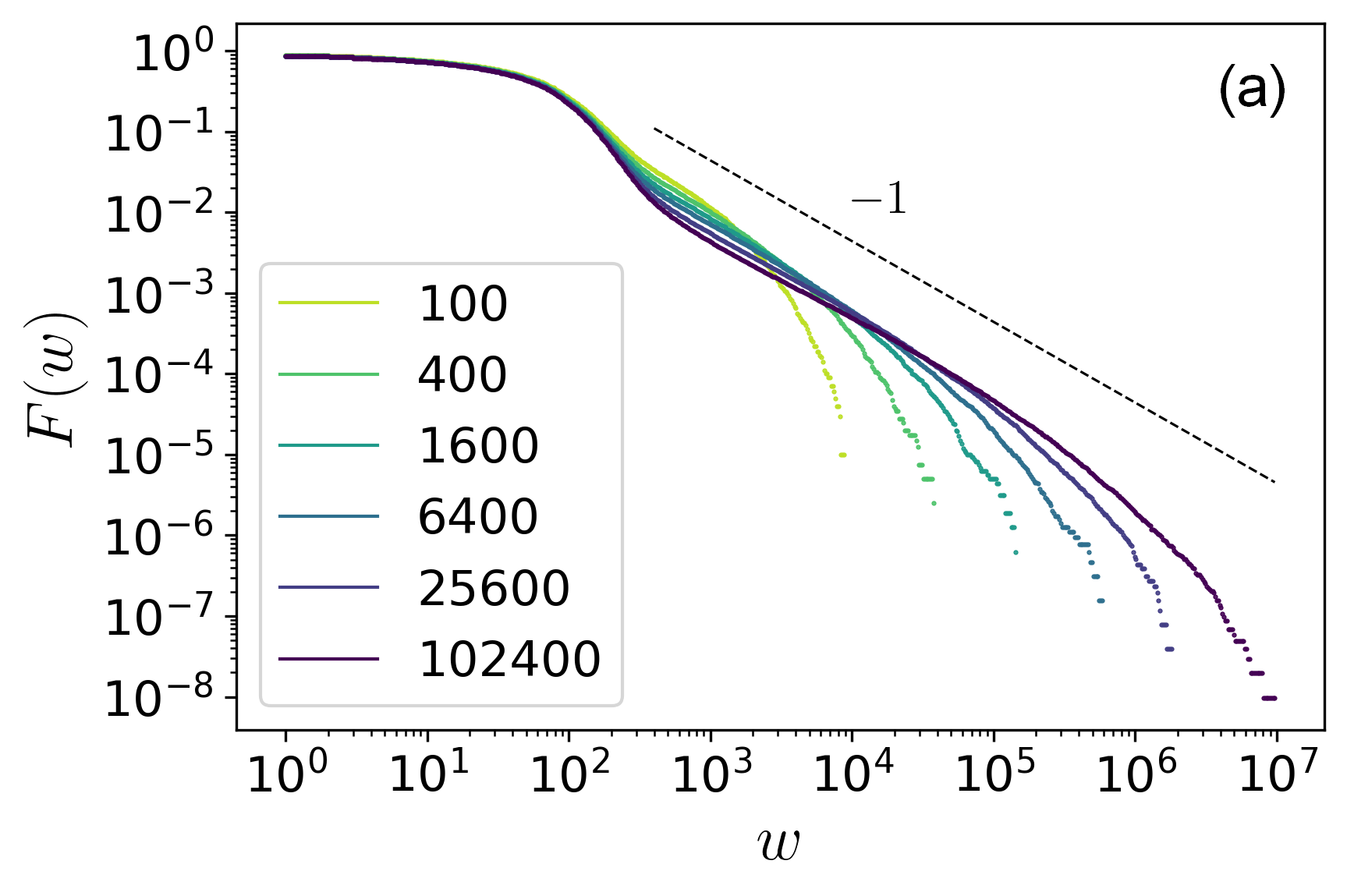}
    \includegraphics[width = 0.45\textwidth]{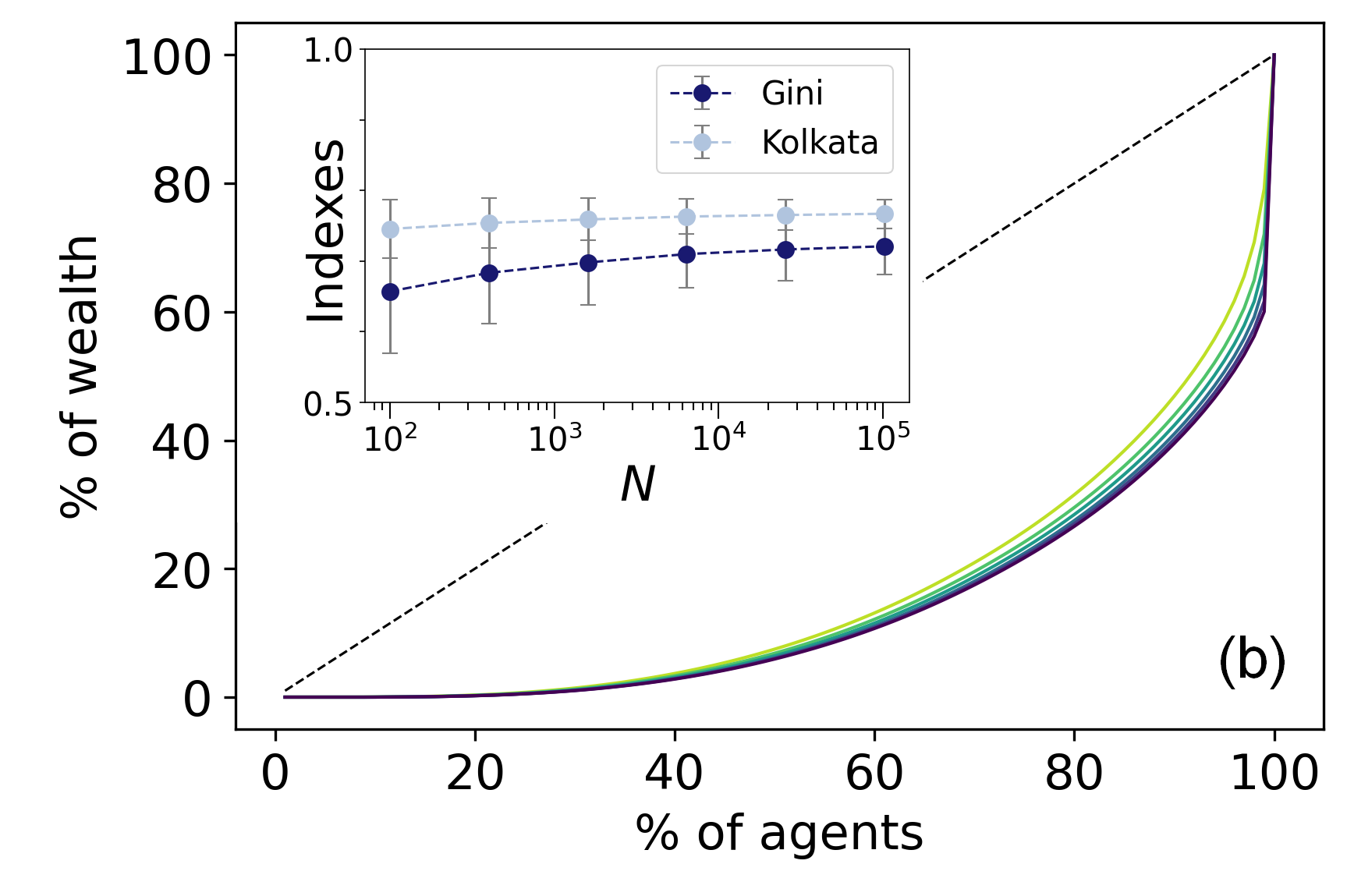}
    \caption{(a) Complementary cumulative  wealth distribution of the full population, $F(w)$,  and (b) corresponding Lorenz curves, for different system sizes $N$ indicated in the legend,  with $\overline{w}=100$ and wage range $[10,90]$.  In (a), the line with slope -1 is drawn for comparison. In (b), the inset shows the  average annual (over 1000 years) Gini  and Kolkata indexes  vs $N$.  
    In the current and subsequent figures, each distribution was built with values recorded at the end of each year, accumulated over the years of a single realization. For the inequality indices, we present the average of the annual values (symbols) and their standard deviation (represented by bars).}
    \label{fig:distribution-size}
\end{figure}

Examining the square root of the variance of the wealth distribution (shown by error bars in Fig.~\ref{fig:N&w}), we find that it decreases with $N$ for employed and unemployed agents, indicating increased homogeneity within these groups. Conversely, the variance among capitalists rises with $N$, likely due to the extended asymptotic tail seen in Fig.~\ref{fig:wealth-histograms}, which suggests a divergent second moment in the limit $N\to \infty$. As $N$ increases keeping $\overline{w}$ constant (thus raising total wealth $W$), the range of capitalist wealth broadens, resulting in greater heterogeneity and an increased likelihood of encountering extremely wealthy agents, as illustrated in Fig.~\ref{fig:distribution-size}(a).
Notice the change in the asymptotic behavior of $F(w)$ as $N$ increases, where the tail decays more slowly, approaching the law $F(w)\sim 1/w$, indicating a greater concentration of large fortunes within an oligarchic group.  This trend is reflected in the Gini index (or Gini coefficient), which increases gradually with $N$ and tends to saturation, 
as shown in the inset of Fig.~\ref{fig:distribution-size}(b). 
The same trend is reflected in the Kolkata index~\cite{BANERJEE2020} also shown in the inset.

\subsection{Influence of the wealth per capita $\overline{w}$}
\label{sec:wealth-total}
 
Figure~\ref{fig:N&w}(b) shows how the distribution of wealth across the different classes (scaled by $\overline{w}$) is affected when changing the wealth per capita $\overline{w}$. The increase of $\overline{w}$ has a strong effect on the distribution of wealth, notability enhancing the  relative wealth of the capitalists  while reducing the  relative  wealth in the working and unemployed classes.

\begin{figure}[h!]  
    \centering
    \includegraphics[width = 0.45\textwidth]{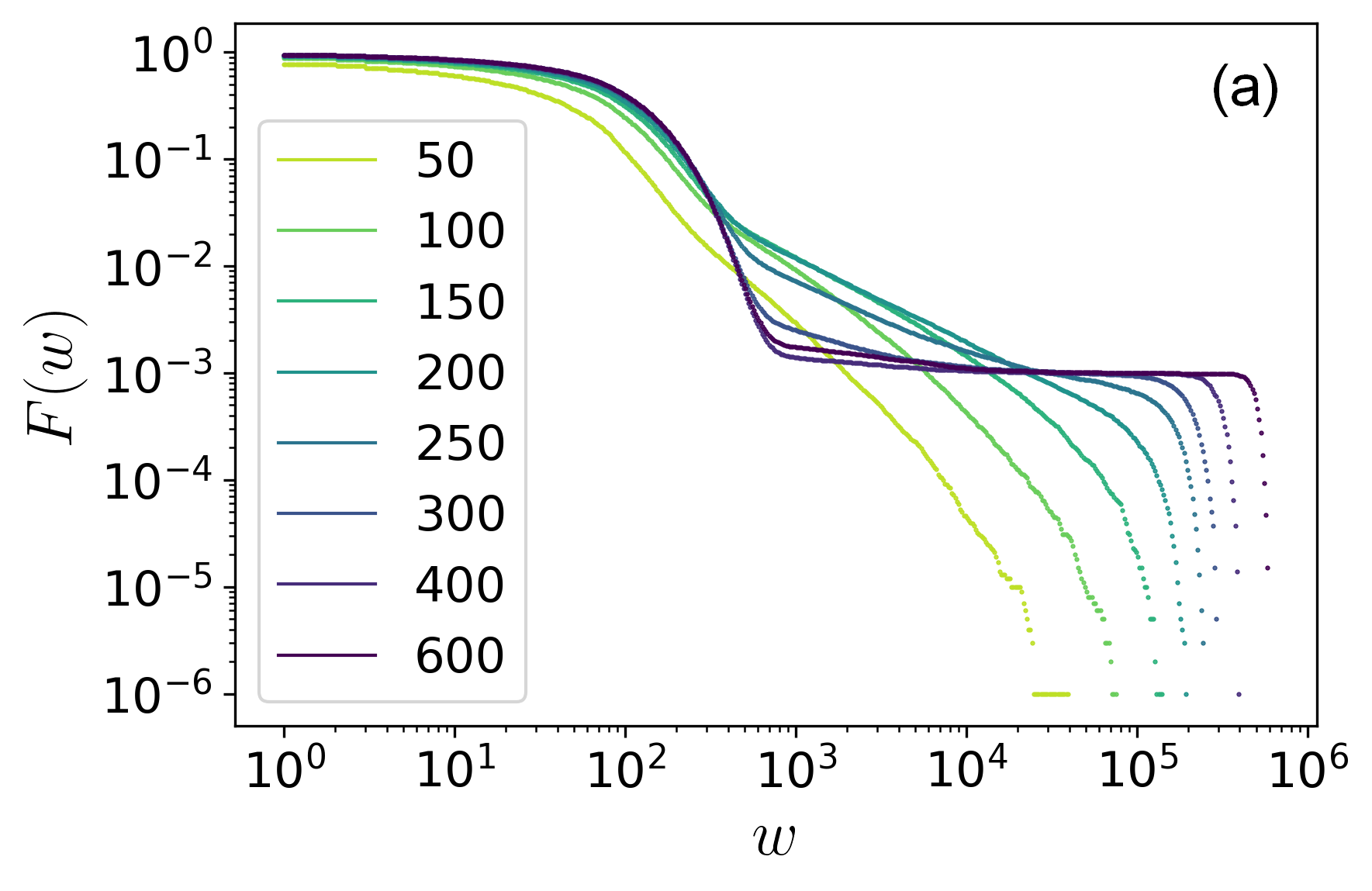}
    \includegraphics[width = 0.45\textwidth]{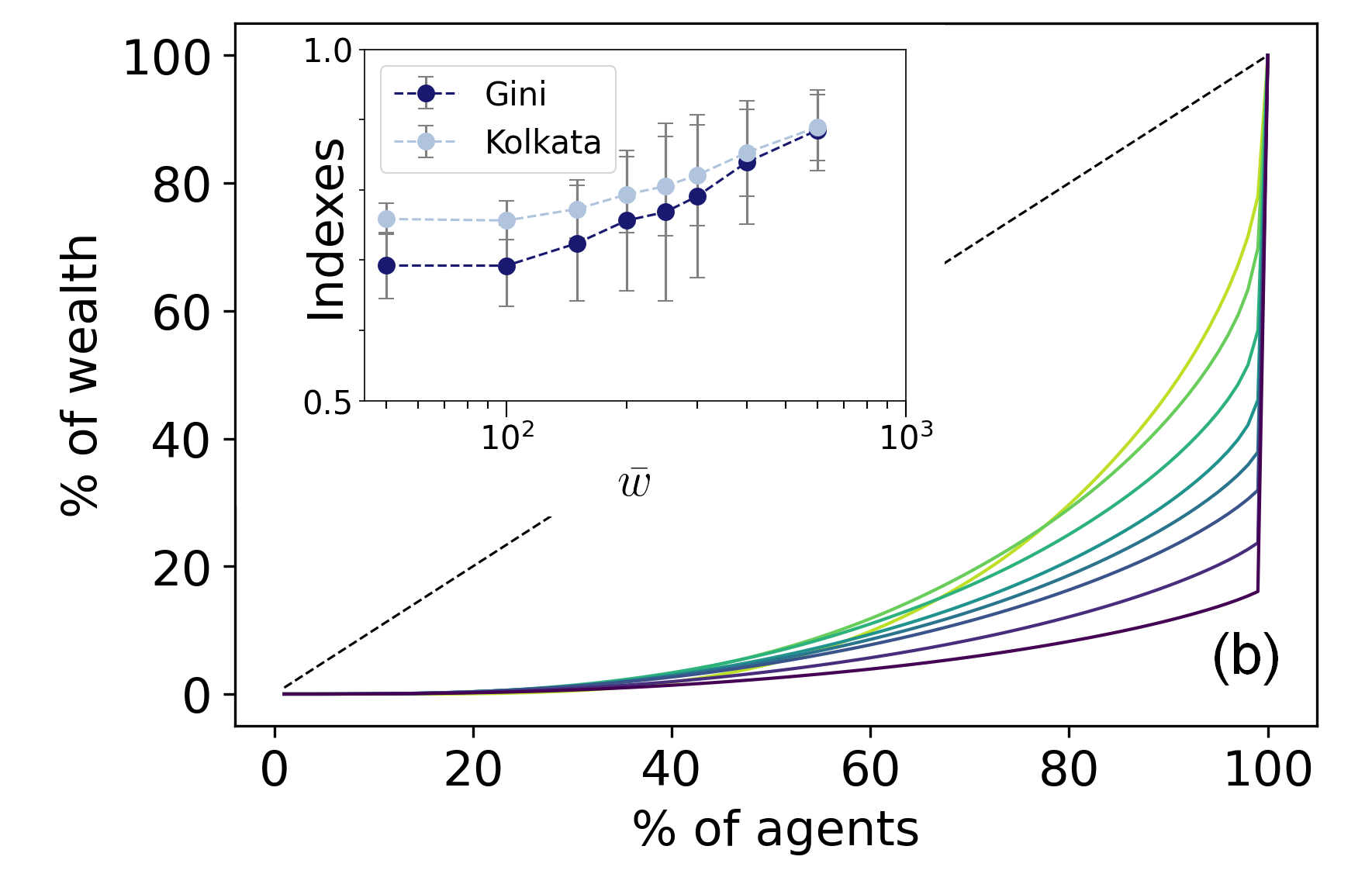}
 \caption{(a) Complementary cumulative  wealth distribution of the full population, $F(w)$, and (b) corresponding Lorenz curves, for different levels of the wealth per capita $\overline{w}$, indicated in the legend. The population size is $N=1000$ and the wage range $[10,90]$.  In (b), the inset shows the  average annual Gini  and Kolkata indexes  vs $\overline{w}$.   
  }
  \label{fig:Fvsw-barw}
\end{figure}

Upon closer examination of the cumulative distribution in Fig.~\ref{fig:Fvsw-barw}(a), it is clear that the tail decays more slowly as $\overline{w}$ increases.
 Recalling that this second regime is associated with the capitalist class, according to Figs.~\ref{fig:portrait} and \ref{fig:wealth-histograms}, the observed flattening implies that capitalist wealth tends to concentrate in a more exclusive oligarchy with the higher $\overline{w}$. 
In other words, the probability of finding agents with very high wealth increases. This is an expected effect, also captured by the Lorenz curves in Fig.~\ref{fig:Fvsw-barw}(b), showing that an increase in total wealth leads to a noticeable increase of the Gini index, i.e., the inequality increases. 
This illustrates the common mistake of using a high gross domestic product as an indicator of progress and equality of opportunities!

For sufficiently large values of $\overline{w}$, (e.g., for $\overline{w}\gtrsim 300$ in the figure), there is a drastic change in the profile of $F(w)$, where the second regime flattens. This indicates the appearance of a very wealthy class,  separated by a gap in the wealth distribution. Moreover, let us remark that, in the extreme of $R \equiv\overline{w}/\overline{p} \gg 1$, the system can freeze with the condensation of the workforce in an oligarchy that concentrates most part of the wealth of the system and their firms cannot bankrupt, which is what happens in the case of the largest value of $\overline{w}$ plotted.

\subsection{Influence of the wage interval}
\label{sec:wage}

\begin{figure}[h!]  
    \centering
\includegraphics[width = 0.45\textwidth]{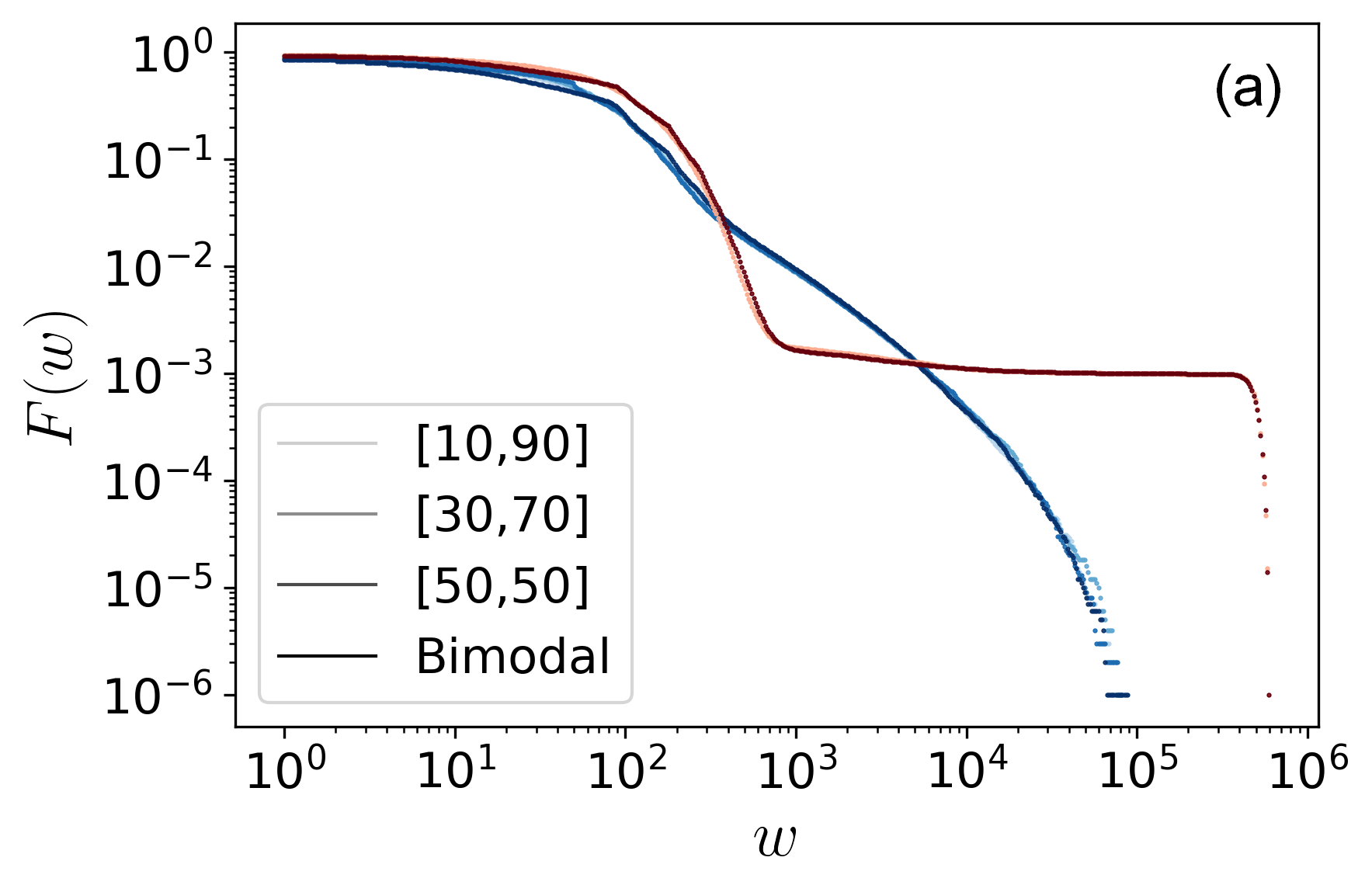}
\includegraphics[width = 0.45\textwidth]{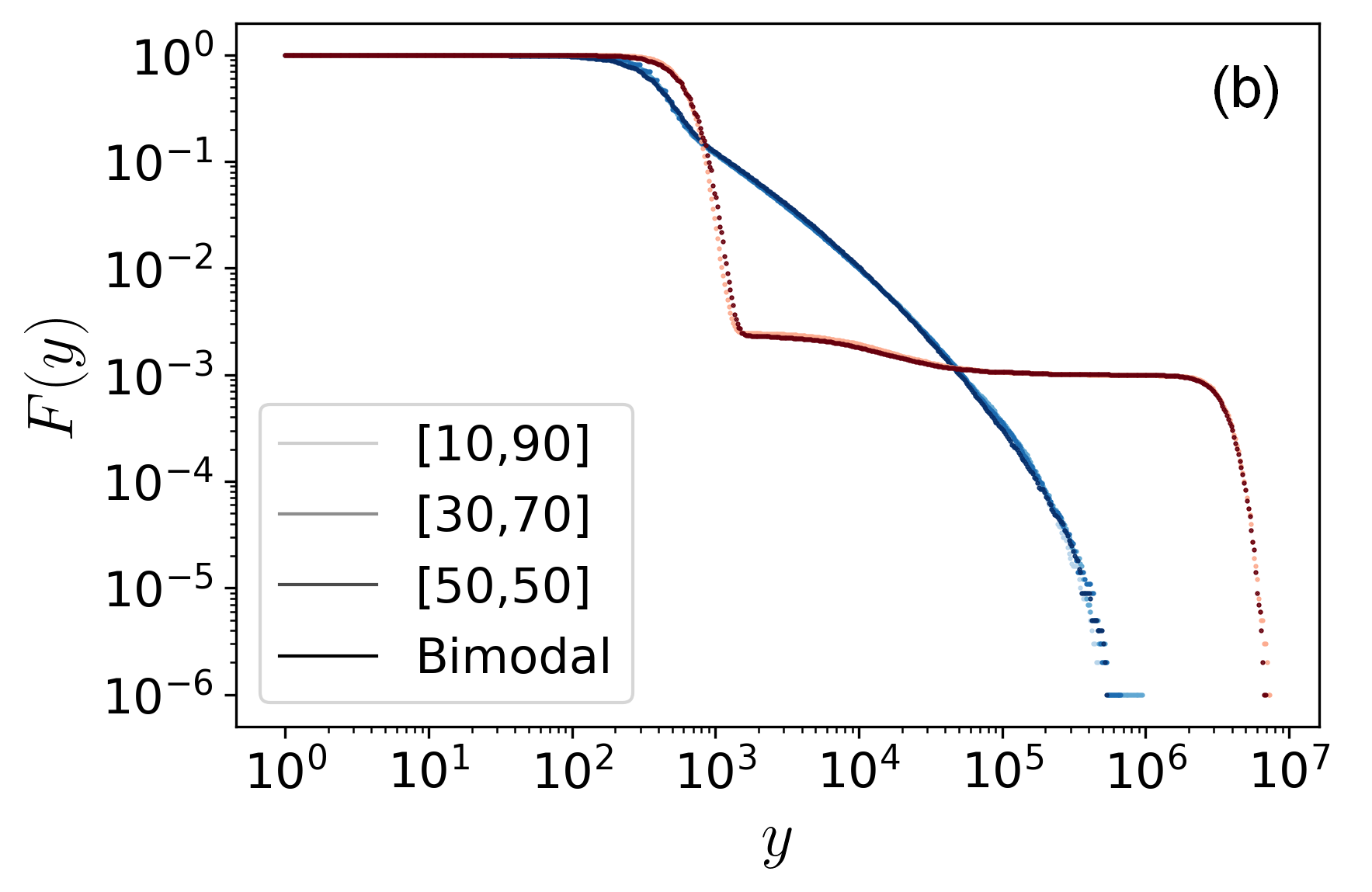}
\caption{ 
Complementary cumulative distributions of (a) wealth, $F(w)$, and (b) income, $F(y)$, for different wage ranges $[p_a,p_b]$ indicated in the legend. The bimodal case corresponds to $p=10$ or $p=90$, with equal probability 1/2. 
$\overline{w}=100$ (blue),   600 (red), and   $N=10^3$.
Note that, in all cases, $\overline{p}=50$. 
}
\label{fig:wageranges}
\end{figure}

We analyze various wage ranges, $[p_a, p_b]$, 
where salaries are uniformly distributed, and explore their impact on the distribution of wealth.
By varying the wage intervals while keeping the mean value $\overline{p}$ fixed, 
we found that the results remain notably invariant to the wage range, as shown in Fig.~\ref{fig:wageranges}(a). 
This is in agreement with the results for uniform distributions reported in Ref.~\cite{lin-lux}. 
The same holds for bimodal distributions, as illustrated in the figure for $p=10$  and 90, with probability 1/2. We also explored non-uniform salary distributions, finding again no significant differences. Consequently, a  significant conclusion not addressed in the original model is that the effect of wages is determined by their mean value $\overline{p}$. 
Furthermore,  the same results are consistently observed for the same ratio $R=\overline{w}/\overline{p}$. Thus, decreasing $\overline{p}$ exerts an identical effect as increasing $\overline{w}$, both enhancing inequality.

\section{Income distribution}
\label{sec:income} 
In this section, we focus on the distribution of incomes (earnings), applying a similar analysis as for the wealth distribution. 
The results for income reveal trends similar to those for wealth in relation to inequality. 
 Let us remark that firm owners (capitalists) collect revenues, while workers receive salaries (wages), and both are computed as incomes.

We consider income to be the sum of all remunerations received by the agent over a one-year period. A worker receives a wage (rule 6), while a capitalist receives a revenue, directly or through its employees (rule 4). Additionally, if an agent changes status within a year, they can receive both types of income: wages and revenues.

\begin{figure}[h!]    
\centering
\includegraphics[width = 0.45\textwidth]{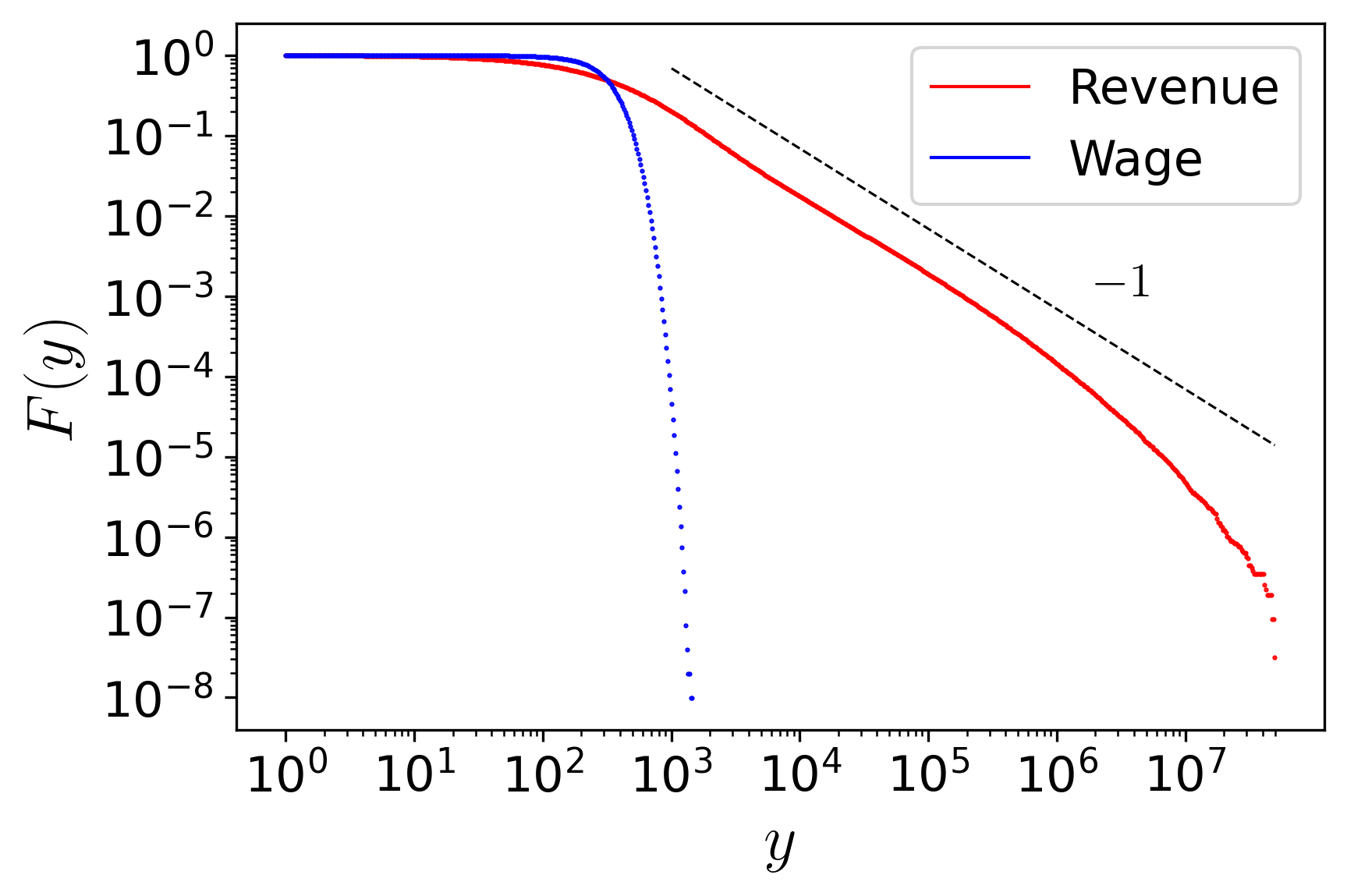}
\caption{Annual income cumulative distribution due to workers' wages and the revenue of firms.  We used $N=102400$,  $\overline{w}=100$, and salary range $[10,90]$. The segment with slope -1 is drawn for comparison.}
\label{fig:renda-fontes}
\end{figure}

The separate distributions of worker's wages and of firms' revenues are shown in Fig.~\ref{fig:renda-fontes}, exhibiting a long tail for revenues. A qualitatively similar result has been previously reported by L. Lin~\cite{lin-lux}, although for other parameter values. Therefore, the distributions of total income,
Fig.~\ref{fig:income-size} (as a function of system size) and Fig.~\ref{fig:income-wealth} (as a function of wealth per capita), both present two regimes that can be associated to each source of income. 
We have shown a similar decomposition before, when discussing wealth, as illustrated in Fig.~\ref{fig:wealth-histograms}.

\subsection{Influence of the system size $N$}

\begin{figure}[h!] 
    \centering
    \includegraphics[width = 0.45\textwidth]{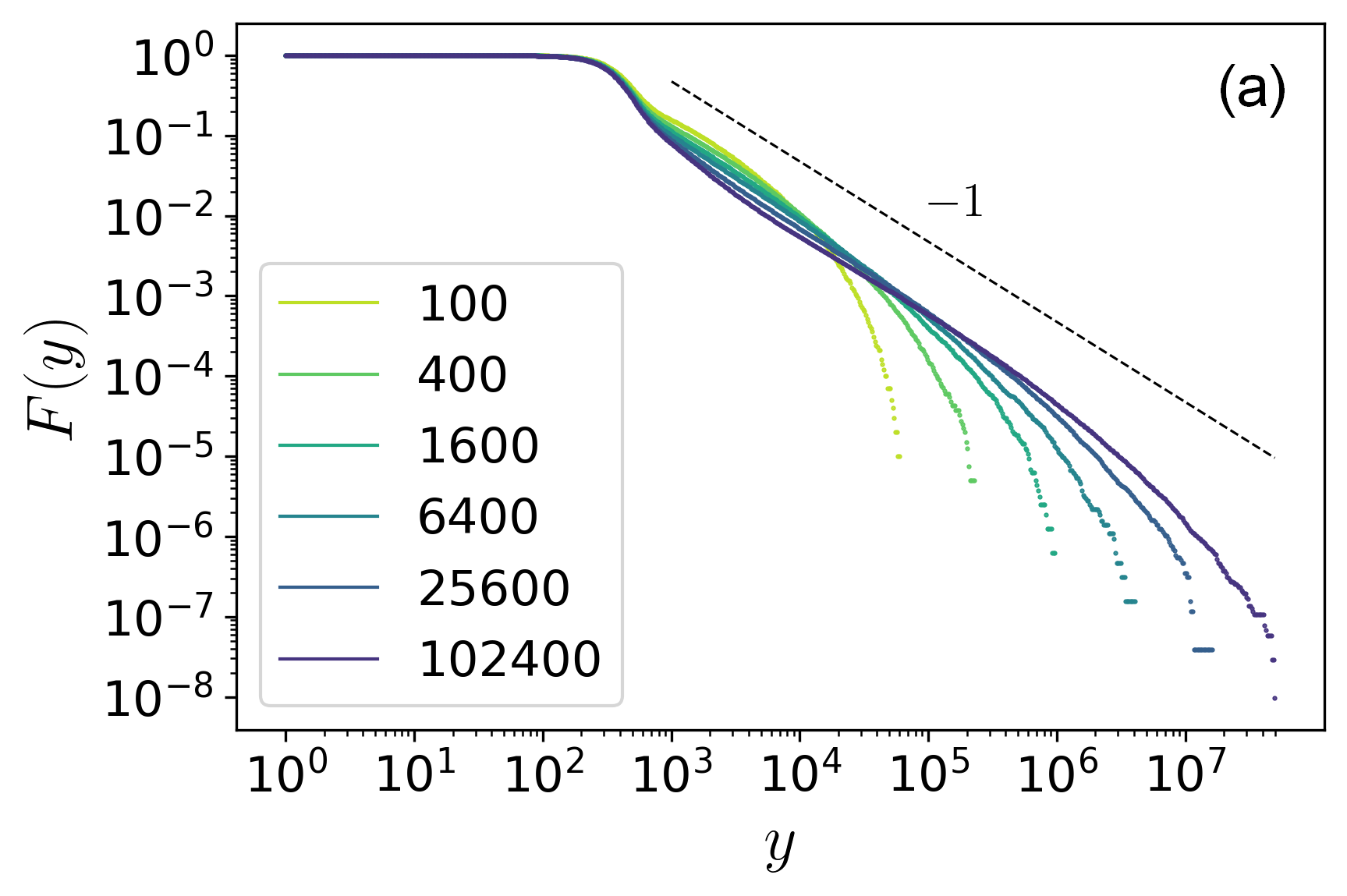}
    \includegraphics[width = 0.45\textwidth]{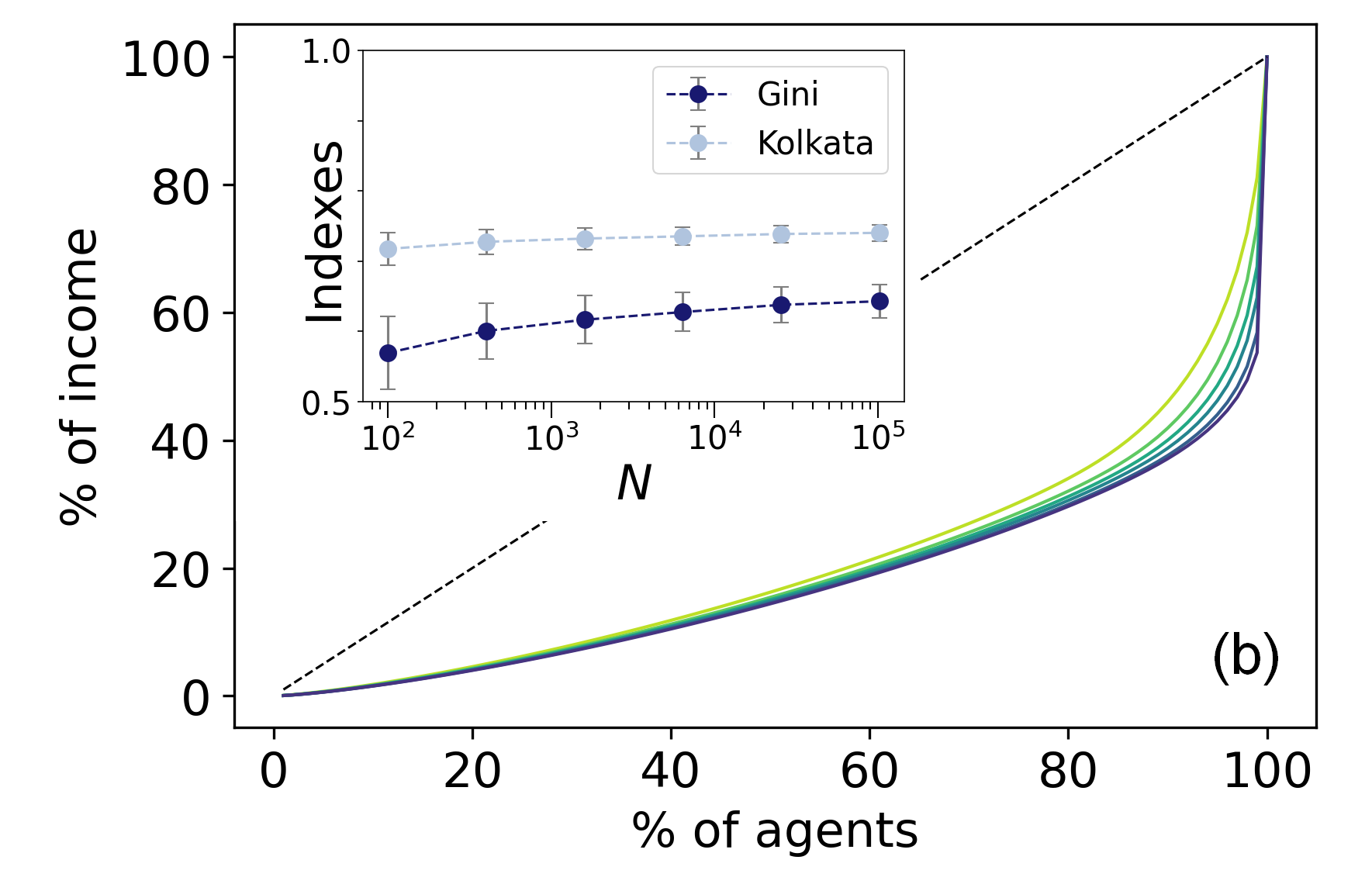}
     \caption{(a)  Annual income complementary cumulative distribution $F(y)$ of the full population and (b) corresponding Lorenz curves, for different system sizes $N$ indicated in the legend,  with $\overline{w}=100$ and wage range $[10,90]$.  In (a), the line with slope -1 is drawn for comparison. In (b), the inset shows the  average annual Gini  and Kolkata indexes  vs $N$.  
    }
    \label{fig:income-size}
\end{figure}

In Fig.~\ref{fig:income-size} we show the impact of the size of the system on the distribution of wealth. 
In panel (a),   we can see how the two-regime distribution evolves as we increase the number of agents, with a Pareto regime similar to that of the wealth distribution.
In the inset of panel (b), we observe a slow increase of the inequality with $N$, as measured by the Gini  and Kolkata indexes, similarly to the behavior of wealth distribution with $N$. 

\subsection{Influence of the  wealth per capita $\overline{w}$}
\begin{figure}[h!] 
    \includegraphics[width = 0.45\textwidth]{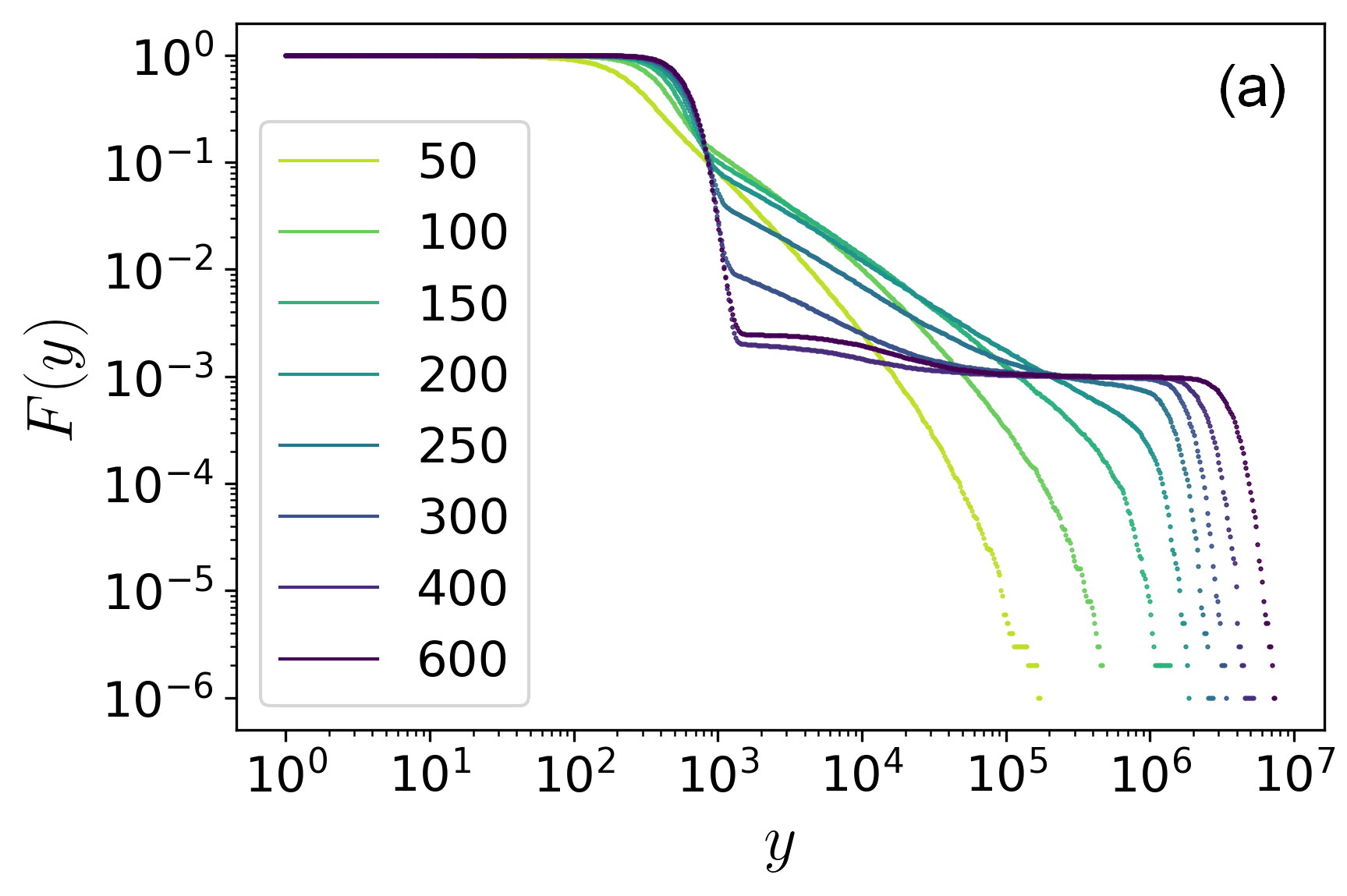}
    \includegraphics[width = 0.45\textwidth]{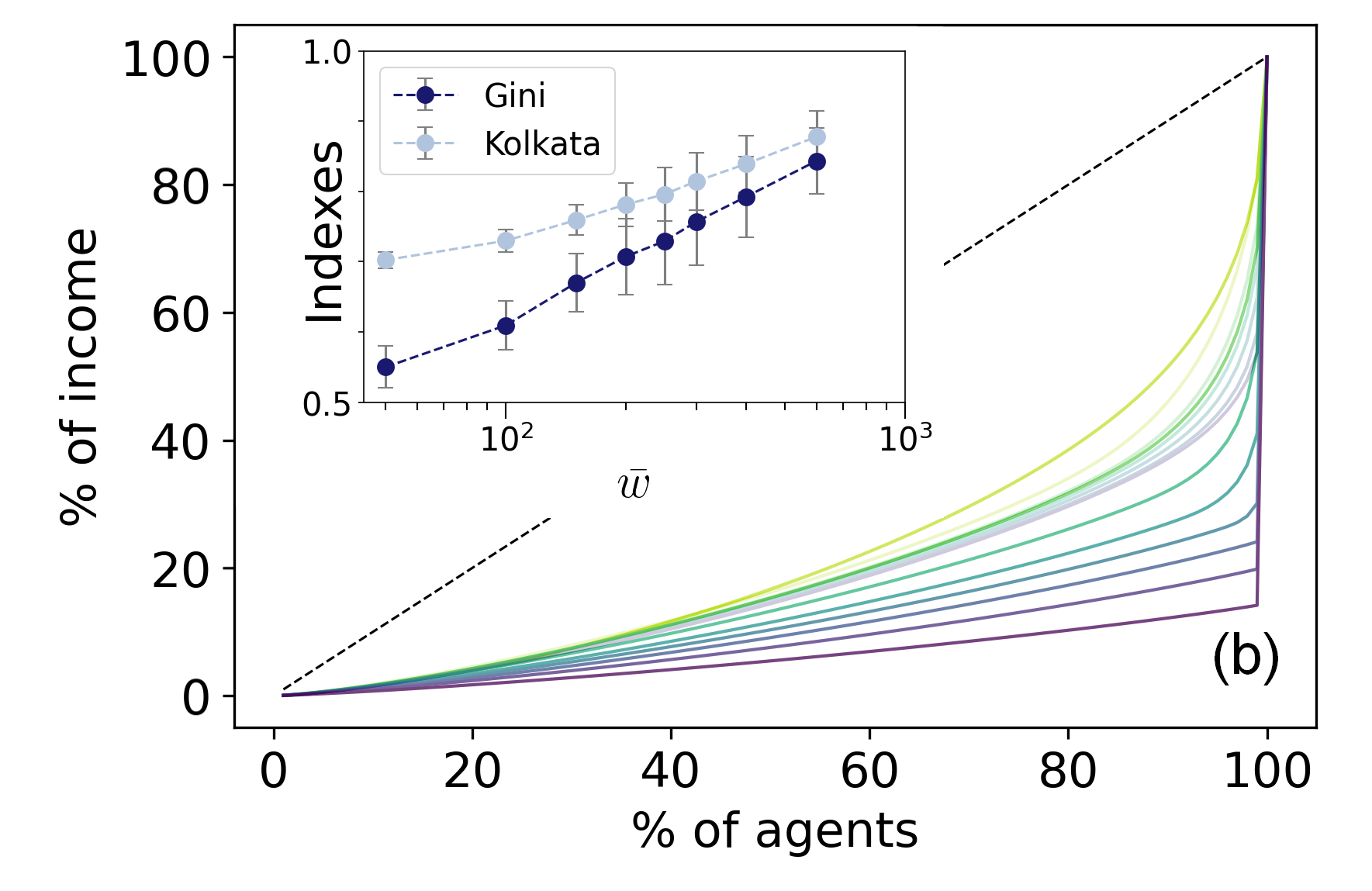}
    \caption{  
    (a) Annual income complementary cumulative distribution $F(y)$ of the full population  $F(y)$, 
    and (b) corresponding Lorenz curves,  for different values of the wealth per capita.  Annual Gini and Kolkata indexes   vs. $\overline{w}$ is in the inset.
The system size is $N=1000$ and the wage range is  [10,90].  \\
    }
    \label{fig:income-wealth}
\end{figure}

Fig.~\ref{fig:income-wealth} shows how the system responds when we increase the  wealth per capita. 
Similarly to the behavior found in the previous section for the wealth distribution,  a strong increase of income inequality also occurs, as measured by the Gini  and Kolkata indexes, shown in the inset of panel (b), which originates from the cumulative wealth distribution and Lorenz curves displayed in Figs.~\ref{fig:income-wealth} (a) and (b), respectively.   
 
\subsection{Influence of  wage interval}

Regarding the wage interval, we confirm here the same behavior observed in wealth distributions, which depends only on the average wage $\overline{p}$, as shown in Fig.~\ref{fig:wageranges}(b).
Therefore,  increasing $\overline{p}$ produces a rapid decay of income inequality.

\section{Final remarks and perspectives}
\label{sec:final}
We have conducted a detailed study of the so-called Social Architecture (SA) model, introduced by Wright in 2005~\cite{WRIGHT2005589}.
The model is based on a few simple rules that govern status changes and wealth exchanges. 
The former occur through hiring and firing, while wealth exchange happens in two ways: directly, as employers pay salaries, and indirectly, as (all) agents contribute to a common fund from which employers retrieve their profits.  In contrast to most models in the literature discussed in Sec.~\ref{sec:introduction}, starting from initial conditions in which agents are in  equal economic and social positions, the the SA model system self-organizes into two classes with a wealth distribution characterized by the two-regime real-world shape. 

Wright presented the model with four parameters: system size $N$, wealth per capita $\overline{w}$, and the lower/upper limits of the salary range, $p_a$ and $p_b$.  
Regarding system size, we have shown that for sufficiently large $N$, the average values of wealth and income within each class become independent of $N$, although long-tails develop with increasing $N$ for the capitalist class. Moreover,  the same qualitative results are observed even for systems as small as $N=100$. 
In terms of the salary range, the results are sensitive only to the average value $\overline p = (p_a+p_b)/2$.
Moreover, since the quantities appearing in the governing rules can be scaled ---indicating a choice of monetary unit without altering the dynamics--- we arrive to the conclusion that the only remaining relevant parameter is the ratio $R=\overline{w}/\overline{p}$. On the one hand, $R\geq1$ is required for the dynamics to evolve from the initial state, as agents would otherwise not have enough wealth to hire employees on average. Furthermore, if $R$ is too large, it leads to an anomalous concentration of wealth and a freezing of the dynamics. Remarkably, within the regime $R\geq1$, this single parameter model yields a two-regime distribution of wealth and income: Boltzmann-Gibbs-like for low values and Pareto-like at the tails.

 To quantify the inequality in wealth and income distributions, we utilized both the Gini and the Kolkata indexes~\cite{BANERJEE2020}. Their dependence on the system size $N$ and on the average wealth $\bar{w}$,  shown in the insets of Figs.~\ref{fig:distribution-size}b, \ref{fig:Fvsw-barw}b, \ref{fig:income-size}b and \ref{fig:income-wealth}b, display similar trends, as expected. 
According to the model's predictions, we have seen that increasing total wealth in the system, while keeping wages fixed, does not alleviate inequality.  On the contrary, it amplifies wealth concentration within the capitalist class and reduces the average wealth of employees, thereby increasing inequality, as evidenced by the wealth distribution and the inequality indexes. Similar effects are observed with respect to income. These results align with the argument that unregulated capitalism drives increasing inequality~\cite{piketty}.
 Since $R$ is the relevant parameter, increasing the average wage has the same effect as decreasing total wealth.

\begin{figure}[h!]  
\centering
\includegraphics[width = 0.45\textwidth]{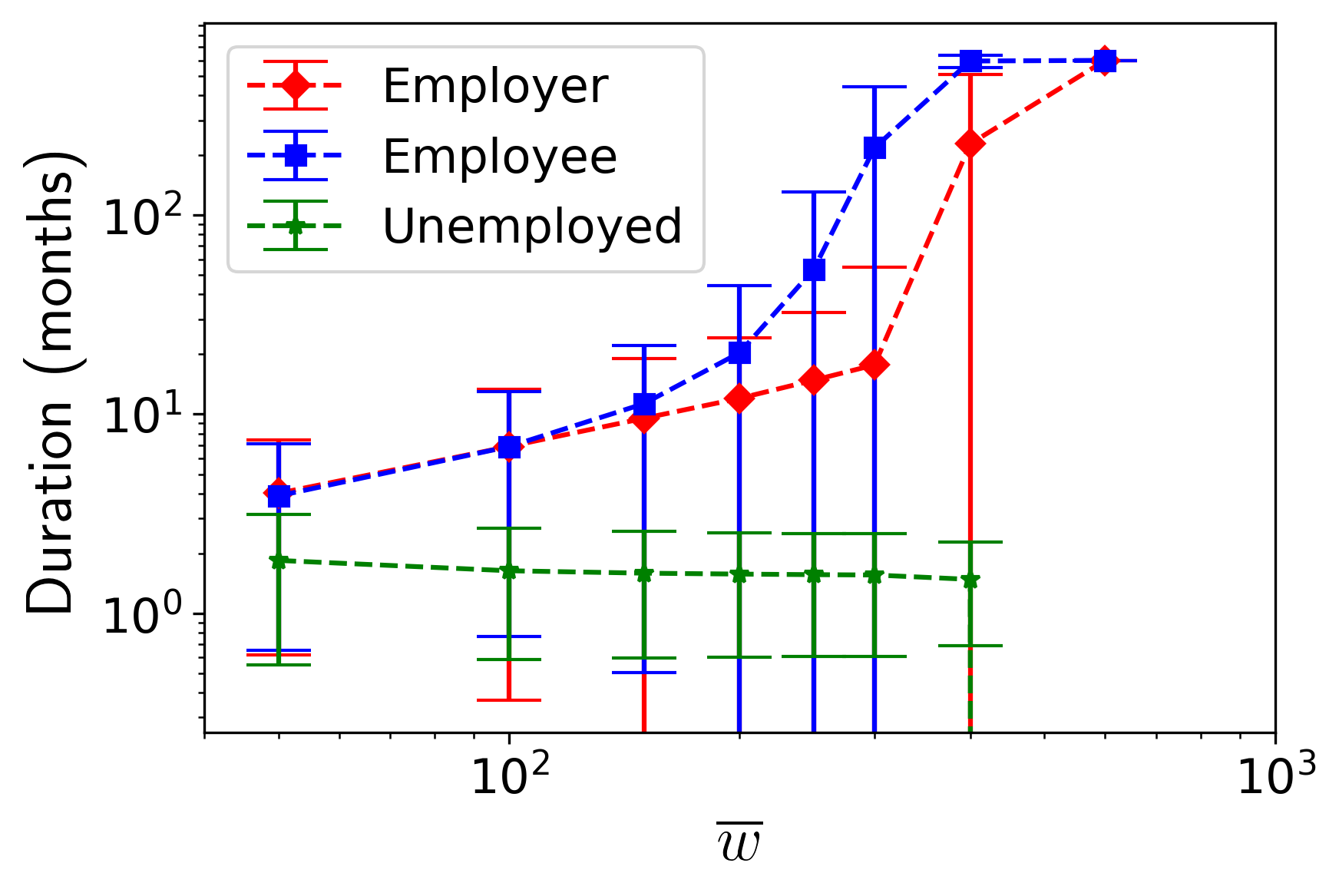}
\caption{Uninterrupted duration (in months) of agents in each class   vs. $\overline{w}$, in 100 simulations up to $t=100$ years (after a transient of 50 years), for $N=1000$ and wage range [10,90]. Symbols represent the average and error bars the standard deviation. }
\label{fig:duration}
\end{figure}

Let us also include and comment an additional simulation result. 
The model yields a high level of mobility between classes, as illustrated in Fig.~\ref{fig:duration}, where we plot the average uninterrupted time each agent spends in a given class as a function of wealth per capita.
For $\overline{w} \simeq 100$,   
the time spent in the worker and capitalist classes is shorter than one year and increases with $\overline{w}$ becoming larger for workers, while the agents remain less time unemployed on average as $\overline{w}$ increases. 
This mobility may seem unrealistic at first sight. 
However, note, for example, that according to data from the US Bureau of Labor Statistics, 50\% of private establishments fail in the first five years~\cite{BLS-US}.
According to the model, this effect can be understood by observing the higher concentration of wealth that makes capitalist agents more robust against decaying to the unemployed class.

 Let us now discuss some model assumptions that could be revised, potentially opening new research perspectives. 
One such assumption is the expenditure rule. In its current form, an agent could theoretically spend their entire wealth in a single transaction. This is clearly extreme when considering a real economy. 
Modifying this rule could help extend the persistence of capitalists within their class. 
In fact, this issue is crucial for employers, as the model makes no distinction between personal and firm wealth, meaning that a business owner could exhaust both in a single transaction! 
In addition, employers hire new employees whenever the opportunity arises, without limitations. It could be interesting to impose constraints related to market value, such as limiting spending to employee savings or firm cash reserves.  Another point is that, in this model, agents are homogeneous and indistinguishable apart from the class to which they belong, with no variation in skill levels. Specifically, hiring assumes that unemployed agents are actively seeking jobs, while employed agents are not pursuing market reallocation, even though people may seek salary upgrades. Some variants addressing this aspect have already been explored~\cite{WRIGHT2010}. 
 
Overall, considering that, as shown in the present work, this is essentially a one-parameter model (namely, $R$),  the results are remarkable. 
Unlike other valuable models which reproduce the Pareto law or even the two regimes mentioned in Section~\ref{sec:introduction}, the SA model stands out for its simplicity. It requires no additional rules or interventions to produce a two-class distribution, offering a fundamental approach for further investigation on the dynamics of wealth distribution and inequality within a simplified economic framework.

{\bf Acknowledgments:}
We acknowledge financial support from CAPES, grant 88887.969088/2024-00 and finance code 001. 
C.A. acknowledges partial financial support from Brazilian agencies CNPq (311435/2020-3) and FAPERJ  (CNE E-26/204.130/2024). S.G. would also like to thank CNPq for partially supporting this work under grant 314738/2021–5.

\bibliography{references}

\end{document}